\documentstyle[12pt,aaspp4]{article}

\begin{document}
\title{On the Nature of the Bursting X-Ray Pulsar GRO J1744-28}
\author{Steven J. Sturner\altaffilmark{1} \& Charles D. Dermer}
\affil{E. O. Hulburt Center for Space Research, Code 7653,\\ Naval Research Laboratory, Washington, DC
20375-5352}
\altaffiltext{1}{NRL/NRC Research Associate}

\begin{abstract} The unusual properties of the bursting X-ray pulsar GRO J1744-28 are explained in terms of a
low-mass X-ray binary system consisting of an evolved stellar companion transferring mass through Roche-lobe
overflow onto a neutron star, implying that the inclination of the system is $\lesssim 18^\circ$.  Interpretation of
the QPO at frequency  $\nu_{\rm QPO} = 40$ Hz using the beat-frequency model of Alpar
\& Shaham  and the measured period derivative $\dot P$ with the Ghosh \& Lamb accretion-torque model implies
that the persistent X-ray luminosity of the source is approximately equal to the Eddington luminosity and that the
neutron star has a surface equatorial magnetic field
$B_{\rm eq}
\cong  2\times 10^{10}\ (\nu_{\rm QPO}/40\ {\rm Hz})^{-1} \ {\rm G}$ for standard neutron star parameters. 
This implies a distance to GRO J1744-28 of
$\cong 5\ (\nu_{\rm QPO}/40\ {\rm Hz})^{1/6} b^{1/2}$ kpc, where $b<1$ is a correction factor that
depends on the orientation of the neutron star.  
\end{abstract}


\section{Introduction}

GRO J1744-28, a 2.1 Hz X-ray pulsar with unusual bursting properties, was discovered on 2 December 1995 in the
direction of the Galactic Center with the Burst and Transient Source Experiment (BATSE) on the {\it
Compton~Gamma~Ray~Observatory}  (Kouveliotou et al.
\markcite{Kouvea96}1996a; Finger et al. \markcite{Fingera96}1996a).  The source initially bursted about every 3
minutes, then declined in frequency to about 30 per day for the first month, after which it increased in frequency  to
about 40 per day in the period 8-18 January 1996 (Fishman et al. \markcite{Fishm96}1996).   The {\it
Rossi~X-ray~Timing~Explorer} ({\it RXTE}) (Swank et al. \markcite{Swank96}1996) localized the source to galactic
coordinates $l= +0.02^\circ, b=+0.3^\circ$.   Both the persistent and bursting X-ray emissions show pulsations with a
0.467 s period (Kouveliotou et al.
\markcite{Kouveb96}1996b) and have similar hard X-ray and soft gamma-ray spectra (Briggs et al.
\markcite{Brigg96}1996; Strickman et al. \markcite{Stric96}1996) with a  color temperature $\approx 10$ keV,
characteristic of X-ray pulsars.  

In this {\it Letter}, we show that the observations are best explained by a neutron star accreting matter through
Roche lobe overflow from an evolved low-mass stellar companion.  We derive values for the magnetic field stength
and accretion luminosity by noting that the beat frequency model (Alpar \& Shaham \markcite{Alpar85}1985) for
QPOs and the accretion-torque model (Ghosh \& Lamb \markcite{Ghosh79}1979) for the pulsar spin-up rate
separately imply  relations between these two quantities.  The size and temperature of the polar-cap hot spot are
determined.  The observed pulse fraction is used to constrain the orientation and geometry of the neutron star, and
to derive the distance to the system from the calculated luminosity.

\section{Nature of the Stellar Companion}

The present mass $M_*=M_*(i)$ of the neutron star's stellar companion can be calculated as a function of orbital
inclination angle $i$ using the measured (Finger, Wilson, \& van Paradijs
\markcite{Fingerb96}1996b) value of the mass function 
\begin{equation} f_{\rm x} = 1.31 \times 10^{-4}\ M_{\odot} = {\left(M_* \sin{i}\right)^3 \over {\left(M_{\rm
NS} + M_*\right)^2}}.
\end{equation}
\noindent The solution to equation (1) is plotted in Figure 1, assuming a neutron star mass $M_{\rm NS} =
1.4M_{1.4}\ M_\odot$ with 
$M_{1.4} = 1$.   We find that $M_*(i) \lesssim 1 M_\odot$ unless $i\lesssim 6^\circ$.  The mean separation
between the neutron star and its companion is given by
\begin{equation} a(i) = a_{\rm x} \sin{i}\ \left[{{M_{\rm NS}+M_*(i)} \over {M_*(i) \sin{i}}}\right],
\end{equation}
\noindent where $a_{\rm x}\sin{i}= 1.12\ R_\odot =7.76\times 10^{10}$ cm (Finger et al.
\markcite{Fingerb96}1996b).   As can be seen in Figure 1, the values of $a(i)$ range from $\sim 25\ R_\odot$ for
$i>5^\circ$ to
$\sim 80\ R_\odot$ for $i=1^\circ$.  

Mass transfer in a low-mass system occurs through Roche-lobe overflow rather than via a stellar wind (e.g.,
Verbunt \markcite{Verbu90}1990). Given
$M_*(i)$ and
$a(i)$, the average radius of the Roche lobe around the stellar companion, 
$R_{\rm RL}(i)$, can be determined from the expressions
\begin{mathletters}
\begin{eqnarray} R_{\rm RL}(i)  & \simeq & a(i) \left\{0.38+0.2\log{\left[M_*(i)/M_{\rm NS}\right]}\right\}\ ;
\hspace{1in}M_*\gtrsim M_{\rm NS},\\
  & \simeq & 0.46\  a(i) \left[{M_*(i) \over {M_{\rm NS}+M_*(i)}}\right]^{1/3}\ ;\hspace{1.4in}M_*\lesssim M_{\rm NS}
\end{eqnarray}
\end{mathletters}
\noindent (e. g. Verbunt \markcite{Verbu90}1990).  $R_{\rm RL}(i)$  is plotted in Figure 1 and varies from
$4\ R_\odot$ at $i=90^\circ$ to $40\ R_\odot$ at $i=1^\circ$.  Such sizes greatly exceed the radii of main
sequence stars with masses
$M_*(i)$, which we also plot in Figure 1 (e.g., Mihalas \& Binney \markcite{Mihal82}1982).   Accretion via Roche lobe
overflow can occur however if the companion star has evolved off the main sequence, which is possible within the
lifetime of the Galaxy for stars with initial masses $\gtrsim 1.0\ M_\odot$.   A  companion with a current mass $>1\
M_\odot$ requires that the system be nearly face-on, with
$i\lesssim 6^\circ$, but mass transfer to the neutron star could have decreased the companion's present mass even if
its initial mass was
$> 1\ M_\odot$.    

The evolved companion star must presently have a radius $R_*\gtrsim R_{\rm RL}(i)$. The radii of evolved
low-mass stars depend mainly on their helium core masses (Webbink, Rappaport, \& Savonije
\markcite{Webbin83}1983). We can use the relations of Webbink et al. \markcite{Webbin83}(1983) to place a lower
limit on the present mass of the companion by requiring it to be larger than the helium core mass.  In the inset
to Figure 1, we plot the stellar radius of an evolved companion versus its helium core mass as well as the Roche
lobe radius versus stellar mass.   We find that the smallest helium core mass which allows for Roche-lobe
overflow in GRO J1744-28 is
$\sim 0.22\ M_\odot$. Thus $M_*\gtrsim 0.22\ M_\odot$ and therefore
$i < 18^\circ$. The chance probablity for viewing the system with this  inclination is 1 in $\sim$20.  Also shown in
Figure 1 is the no-eclipse condition assuming that $R_* = R_{\rm RL}$.  This restricts $i\lesssim 80^\circ$, which is
less constraining than the above considerations.

\section{Accretion Dynamics}

The Alfv\'en radius for spherical accretion, obtained by balancing accretion and magnetic pressure, is given by
\begin{equation} R_{\rm A }= 1.9\times 10^8\  \mu_{30}^{4/7}\ M_{1.4}^{1/7}\ (L_{38}\ R_6)^{-2/7}\ \ {\rm
cm},\end{equation}
\noindent where $L_{38}$ is the pulsar luminosity in units of $10^{38}\ {\rm erg\ s^{-1}}$, $\mu = B_{\rm
eq}R_{\rm NS}^3 = 10^{30}\mu_{30}$ G-cm$^3 $ is the neutron star magnetic moment ($B_{\rm eq}$ is the
surface equatorial field strength), and
$10^6R_6$ cm is the neutron star radius. The corotation radius, where the Keplerian and rigid body rotation
velocities are equal, is given by $R_{\rm co} = (GM_{\rm NS})^{1/3} (P/2\pi)^{2/3} = 1.0\times 10^8\
M_{1.4}^{1/3}$ cm using
$P =  0.467$ s for GRO J1744-28. 
The transition from Keplerian motion to corotation with the
neutron star magnetosphere occurs at $r = \eta R_{\rm A}$, where $\eta \approx 1$.  The requirement
that $\eta R_{\rm A} < R_{\rm co}$ for acretion to take place yields $ B_{\rm eq} \lesssim 3.3\times 10^{11}\ L_{38}^{1/2}\
M_{1.4}^{1/3}\ R_6^{-5/2}\eta^{-7/4}\  {\rm G}$. Requiring magnetic confinement of the hot plasma in the accretion column sets a
lower limit on the surface polar magnetic field strength $B_{\rm p}$ ($=2B_{\rm eq}$ for a dipole magnetic field) 
through the expression
$B^2_{\rm p}/8\pi \gtrsim L/A_{\rm pc} c \approx \sigma_{\rm SB}T^4/c$, where we use the blackbody radiation
formula to estimate the polar cap area $A_{\rm pc}$.  Thus $B_{\rm p} \gtrsim 2.9\times 10^9\ T_{10}^{2}\  {\rm G}$,
where $T_{10}$ is the effective temperature in units of 10 keV.

A more precise estimate for magnetic field strength and accretion luminosity can be obtained by interpreting the
measured QPO frequency using the beat frequency model of Alpar \& Shaham \markcite{Alpar85}(1985), and the 
pulsar spin-up rate from the Ghosh \& Lamb
\markcite{Ghosh79}(1979) accretion-torque model.  Zhang et al. \markcite{Zhange96}(1996) report a QPO at
frequency $\nu_{\rm QPO} = 40$ Hz with integrated power  (uncorrected for deadtime) under the 27 Hz FWHM
Lorentzian equal to 5.6\% of the total power. Although  the width of this QPO is somewhat larger than those
typically measured in low mass X-ray binaries, with FWHM widths $\lesssim 0.5$ the centroid frequencies (see, e.g., van der
Klis
\markcite{Klis95}1995), we still assume that
$\nu_{\rm QPO}$ can be interpreted as the beat frequency between the Keplerian frequency 
$\nu_{\rm K} = (GM_{\rm NS})^{1/2}(2\pi)^{-1} r^{-3/2}$ at  $r =
\eta R_{\rm A}$ and the neutron star's rotation frequency
$\nu_{\rm NS}$ (Alpar \& Shaham \markcite{Alpar85}1985). Less significant peaks in the power spectrum are also found
at 20 and 60 Hz, and we cannot exclude that the 40 Hz peak is a harmonic of the 20 Hz peak but with larger
amplitude.   This implies
\begin{equation} B_{\rm eq} = 1.1\times 10^{10}\ \tilde\nu^{-7/6}\ L_{38}^{1/2}\ M_{1.4}^{1/3}\ R_6^{-5/2}\
\eta^{-7/4}\ {\rm G}
\end{equation}
\noindent and
\begin{equation}
\eta R_{\rm A} = 1.4\times 10^{7}\ \tilde\nu^{-2/3}\ M_{1.4}^{1/3}\ {\rm cm} ,
\end{equation}
\noindent where we define $\tilde\nu = (\nu_{\rm QPO}+\nu_{\rm NS})/40~{\rm Hz}$.

An expression for the period derivative of a disk-fed X-ray pulsar as a function of neutron star properties was derived
by Ghosh \& Lamb \markcite{Ghosh79}(1979).  Substituting equation (5) into their equation (15) gives
\begin{equation} -\dot{P} = 2.6\times 10^{-5}\ \tilde\nu^{-1/3}\ L_{38}\ M_{1.4}^{-1/3}\ R_6\ I_{45}^{-1}\
\eta^{-1/2}\ f(\omega_{\rm s})\ \ {\rm s\ yr^{-1}},
\end{equation}
\noindent where $I_{45}$ is the neutron star moment of inertia in units of $10^{45}\ {\rm g}$-${\rm cm^2}$, and
$f(\omega_s)$, proportional to the dimensionless torque function (see Ghosh \& Lamb \markcite{Ghosh79}1979; Wang
\markcite{Wang95}1995), is a function of the fastness parameter
$\omega_s = (\eta R_A/R_{\rm co})^{3/2}$. For GRO J1744-28, $\omega_{\rm s} \cong 0.05\ \tilde\nu^{-1}\ \ll1$, implying
$f(\omega_s)\approx 0.9$.
  The observed period derivative is  $\dot P = -5.95\times 10^{-5}\ {\rm s\ yr^{-1}}$  (Finger et al.
\markcite{Fingerb96}1996b), which could however be different than the value at the epoch 18-19 January 1996 when {\it
RXTE} measured the 40 Hz QPO.  In terms of the Eddington luminosity $L_{\rm Edd} = 1.76\times 10^{38}\ M_{1.4}\ {\rm
erg\ s^{-1}}$,  we obtain a pulsar luminosity given by 
\begin{equation} L/ L_{\rm Edd} \cong 1.4\ \eta^{1/2}\ \tilde \nu ^{1/3}\
M_{1.4}^{-2/3}\ R_6^{-1}\ I_{45}\ .
\end{equation}
\noindent Substituting equation (8) into (5), we obtain 
\begin{equation} B_{\rm eq} \cong 1.8\times 10^{10}\ \eta^{-3/2}\ \tilde\nu^{-1}\ M_{1.4}^{1/2}\
R_6^{-3}\ I_{45}^{1/2}\ {\rm G}\ .
\end{equation}
\noindent   The derived value for the magnetic field at the neutron star's equator depends only upon quantities
intrinsic to the neutron star and details of the accretion process, and is bracketed by our estimates given above. We
suggest a search with {\it RXTE} or {\it ASCA} for cyclotron absorption features with a fundamental energy at $\sim
1$-$2$ keV; note that a larger field is inferred if one assumes that the 20 Hz peak in the power spectrum is
the fundamental.  The magnetic field given by equation (9) does not conflict with OSSE observations, which show no
evidence for  cyclotron features at $> 40$ keV (Strickman et al.
\markcite{Stric96}1996).

\section {X-Ray Pulsar Geometry}

Equation (6) for the stopping radius yields a size of the polar-cap hot spot if we assume that its area is determined
by the footprints of those magnetic field lines that cross the magnetic equator at $r=\eta R_{\rm A}$.  For an
assumed dipolar magnetic field geometry, $\sin^2
\theta_{\rm m}/r = $ constant, where $\theta_{\rm m}$ is the magnetic colatitude.  Thus for an aligned rotator $\theta_{\rm pc}
=\arcsin\left[0.26\ \tilde \nu^{1/3}\ M_{1.4}^{-1/6}\ R_6^{1/2}\right] \cong 15^\circ$, although the hot spot size of an
oblique rotator could be smaller if the matter does not accrete uniformly onto the polar cap.  The hot-spot size and
source luminosity from equation (8) imply an effective temperature of
$\cong 4.9\ \eta^{1/8}\tilde\nu^{1/12} M_{1.4}^{1/12} R_6^{-3/4} I_{45}^{1/4}$ keV.  This is in
reasonable agreement with the values of 6 and 14 keV for the blackbody temperature and e-folding energy reported by
BATSE (Briggs et al. \markcite{Brigg96}1996) and {\it RXTE} (Swank et al. \markcite{Swank96}1996), respectively.

BATSE reports a pulse fraction in the persistent emission of 40\% in the 25-45 keV band (Finger et al.
\markcite{Fingera96}1996a), whereas smaller energy-dependent values are measured with {\it RXTE} for the
band between 3 and 12 keV (Zhang
\markcite{Zhang96}1996).  These pulsations are probably due to projection effects of the X-ray hot spots on the
rotating neutron star.  Figure 2 shows the various orientations and obliquenesses that  produce pulse fractions of 20,
30, and 40\% calculated from a purely geometric model consisting of two identical uniformly-emitting polar
caps. The large luminosities radiated from a small area may produce a fan-beam
radiation pattern, but are not considered here.  We include shadowing effects from the neutron star but
not from the accretion column.  A large range of angles are permitted for a given pulse fraction. During the bursting
episodes, the flux increases by a factor of 2-8 whereas the spectrum remains nearly identical (a relative excess below
3-4 keV is seen during outbursts by {\it RXTE}; Zhang
\markcite{Zhang96}1996). Interpreting the increase in flux as due to an increase in emitting area,  we predict that
the pulse fraction should decrease by an amount dependent on $\alpha$ (see inset to Figure 2).  The
energy-dependence of the pulse fraction could be a consequence of thermal stratification, with the hotter inner
regions therefore showing a larger pulse fraction.

The 2-100 keV energy flux reported by {\it RXTE} at epoch 18-19 January 1996 (Swank et al.
\markcite{Swank96}1996) is
$\phi = 2\times 10^{-7}$ ergs cm$^{-2}$ s$^{-1}$ for the persistent emission.  The luminosity from equation (8)
then implies a distance to GRO J1744-28 of 
\begin{equation} d = 4.6 \ \eta^{1/4}\ \tilde \nu^{1/6}\ M_{1.4}^{1/6}\ I_{45}^{1/2}\ b^{1/2}\
R_6^{-1/2}\ {\rm kpc},
\end{equation}
\noindent where the geometric factor $b$ depends on neutron-star parameters.  By approximating  the X-ray hot
spot as a flat disk, we find $b
\cong
\cos\theta_{\rm obs}\cos\alpha$ in the limit $\theta_{\rm obs}+\theta_{\rm pc}+\alpha \ll \pi/2$  where 
neutron shadowing effects are unimportant.  For 2-3 magnitudes of extinction per kpc in the direction of the
Galactic Center, we estimate that the brightest optical counterpart would be at $V \cong 20-24$ for an M0 III
giant ($M_V=-$0.4). 

\section {Discussion}

GRO J1744-28 shows characteristics of both X-ray bursters and X-ray pulsars.  Its intermediate magnetic field
strength between the weak fields ($B \approx 10^{8-9}$ G) thought to reside in
low-mass X-ray binaries and the strong fields ($B \approx 10^{12-13}$ G) found in X-ray pulsars suggests that its burst
properties are related to those of X-ray bursters.  But the relative fluences in the burst and persistent emissions  probably
rule out a thermonuclear origin in the initial bursting phase (Kouveliotou et al. \markcite{Kouvea96}1996a; Strickman et al.
\markcite{Stric96}1996), although the energetics derived from {\it RXTE} observations (Swank et al.
\markcite{Swank96}1996) are favorable to such an interpretation.  An accretion-disk instability seems implied by the
similarity of the spectral shapes of the burst and persistent emission, since approximately equal energy per particle
will be dissipated for an accretion-powered burst.  We also note the similarities of the burst precursor and
post-burst recovery period between the bursts from GRO J1744-28 and the Type II bursts of the rapid burster (compare
Dotani et al.
\markcite{Dotan90}1990; Lubin et al.
\markcite{Lubin92}1992), the latter of which are thought to be due to accretion instabilities.

The beat-frequency model implies a testable relation between the persistent luminosity and the QPO centroid frequency
given by equation (8). The inferred persistent luminosity implies a super-Eddington
flux for the bursting episodes, since the burst flux is
$\approx 2-8$ times as great as the persistent emission (Swank et al. \markcite{Swank96}1996). The 3-5 minute
recovery period following a burst observed by {\it RXTE} could be due to radiation effects on the disk.

The onset of detectable X-ray emission from this source cannot be due to orbital effects as in the case of Be X-ray
binaries (the system is nearly circular with eccentricity
$e < 0.026$; Finger et al. \markcite{Fingerb}1996b).  We speculate that as the evolved companion expands and overflows its
Roche lobe, the increased X-ray illumination drives an enhanced fueling episode (see Tavani \& London \markcite{Tavan93}1993)
which ceases when the heated atmosphere no longer overflows the Roche surface.  The star then contracts to a quiescent
phase after which it resumes its slow expansion leading to the onset of another episode of mass transfer and
luminous X-ray emission.  Further calculations of evolution, accretion, and radiation will be required to yield a more
complete understanding of this extraordinary system.

\section {Acknowledgements}  We thank Mark Finger, Eric Grove, Mike Harris, Chryssa Kouveliotou, Jim Kurfess, Gerry Share, Mark
Strickman, Jean Swank, Marco Tavani, Bob Wilson, and Will Zhang for very useful discussions.  This work was partially
supported by the {\it Compton Gamma Ray Observatory} Guest Investigator Program GRO-95-135.

\newpage
\begin{figure}[h]
\plotone{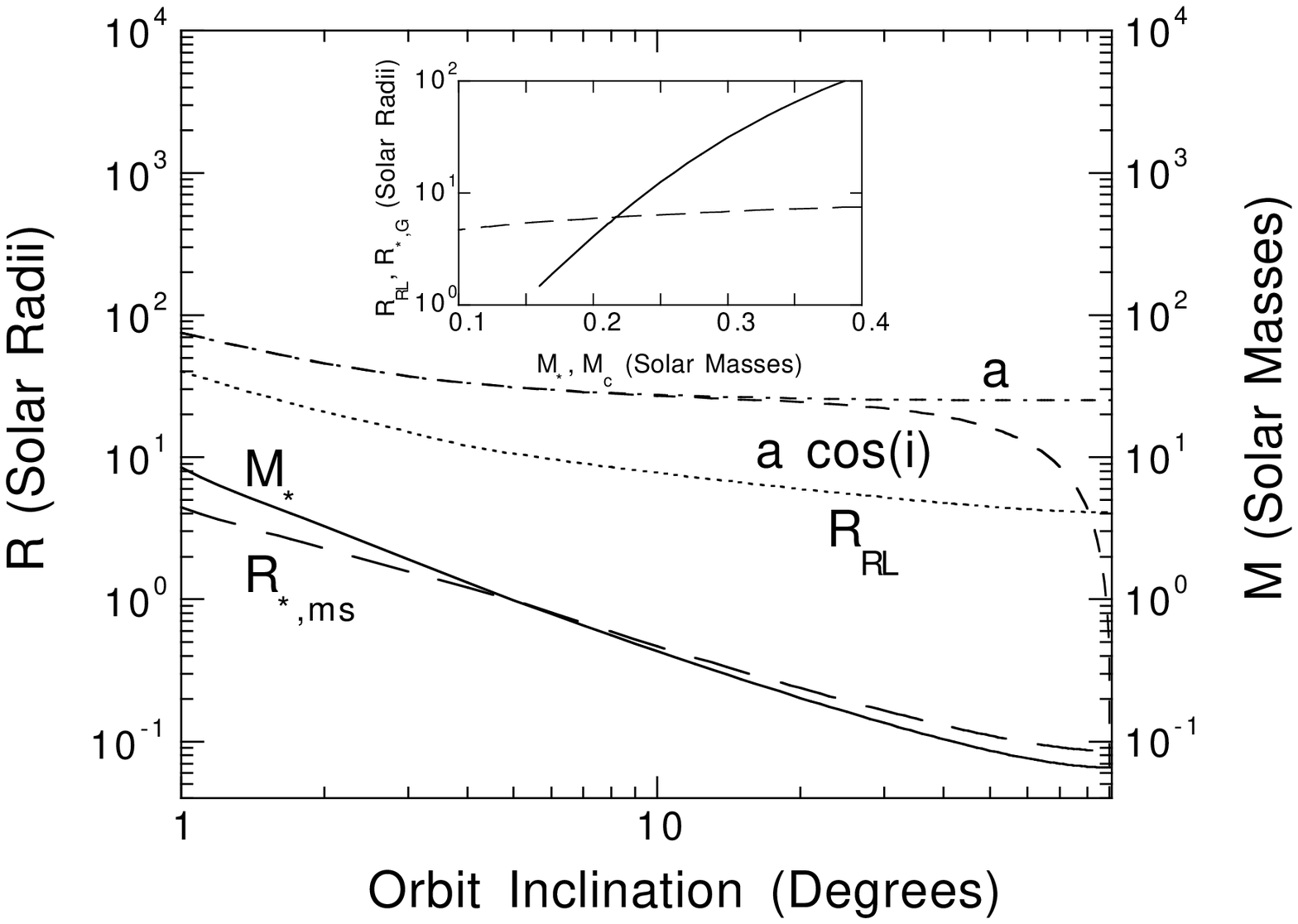}
\caption{Relations between stellar masses and radii which are consistent with orbital parameters of GRO
J1744-28.  We plot the mean separation
$a$, the mass of the stellar companion $M_*$, the main sequence radius $R_{*, {\rm ms}}$ of a stellar companion of
mass $M_*$, and the average Roche lobe radius $R_{\rm RL}$ as a function of the
orbital inclination angle $i$, assuming a 1.4 $M_\odot$ neutron star.    If the companion
fills its Roche lobe, the no-eclipse condition requires $a\ \cos(i)> R_{\rm RL}$.   The inset shows the Roche lobe
radius  $R_{\rm RL}$ (dashed curve) as a function of $M_*$ and the radius $R_{*,g}$ (solid curve) of an evolved
companion as function of its helium core mass $M_{\rm c}$.}  
\end{figure}

\newpage

\begin{figure}[h]
\plotone{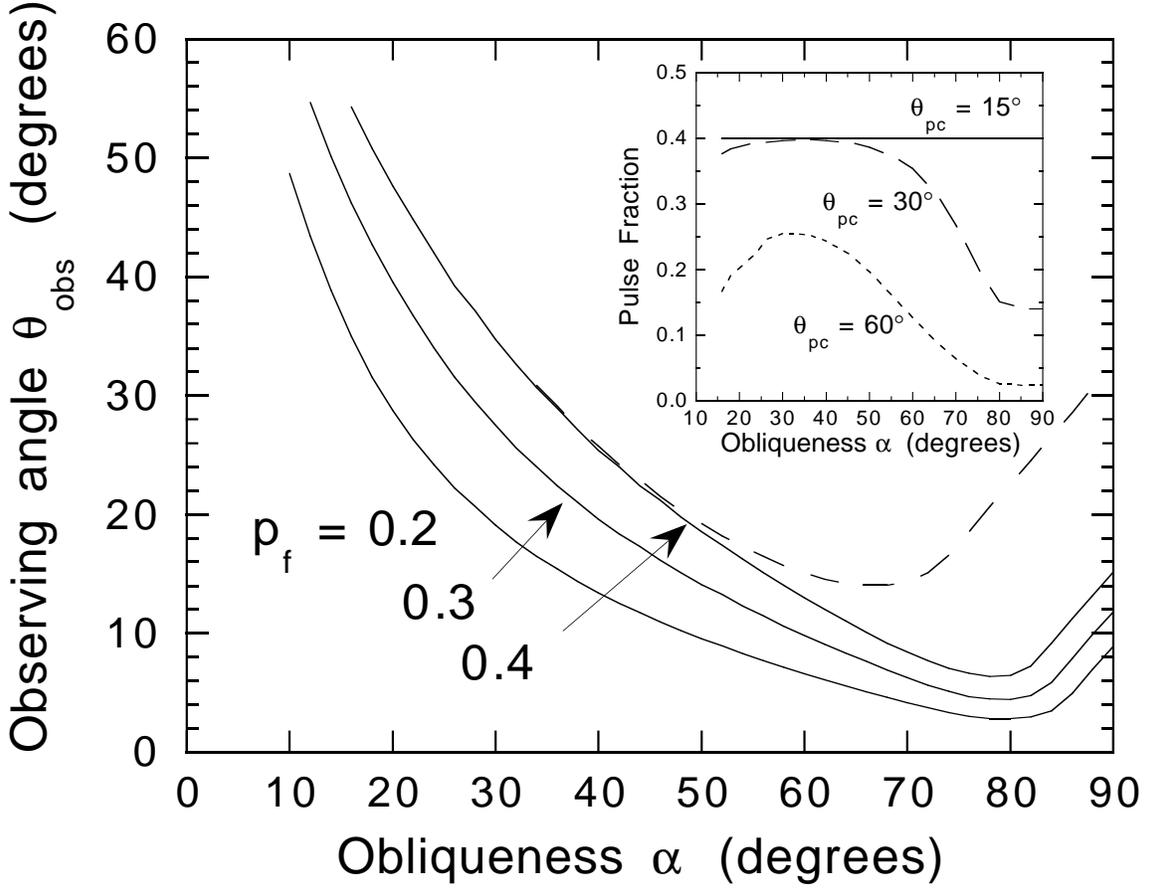}
\caption{Values of $\alpha$ and $\theta_{\rm obs}$ giving pulse fractions 
 $p_f=0.2$, 0.3, and 0.4 for X-ray emitting hot spots with half-angle $\theta_{\rm pc} = 15^\circ$
 (solid curves) and $30^\circ$ (dashed curve).  The obliqueness $\alpha$ is the angle between the
 rotation and magnetic dipole axes, and $\theta_{obs}$ is the angle between
the rotation axis and the direction to the line-of-sight.  We define $p_f = (\phi_{\rm max}- \phi_{\rm
min})/(\phi_{\rm max}+\phi_{\rm min})$, where $\phi_{\rm max(min)}$
 represents the maximum (minimum) fluxes in the pulse profile.   Inset shows variation of
$p_f$ for polar cap sizes $\theta_{\rm pc} = 30^\circ$ and $60^\circ$,
 using values of $\alpha$ and $\theta_{\rm obs}$ which yield a 40\% pulse fraction  when $\theta_{\rm pc} = 15^\circ$.}
\end{figure}

\end{document}